\documentclass[usenames,dvipsnames,sigconf]{acmart}

\usepackage[nolist]{acronym}
\AtBeginDocument{%
 \providecommand\BibTeX{{%
    \normalfont B\kern-0.5em{\scshape i\kern-0.25em b}\kern-0.8em\TeX}}}

\usepackage{soul}
\usepackage{color}
\usepackage{wallpaper}
\usepackage{graphicx}
\usepackage{cleveref}
\usepackage{numprint}
\usepackage{paralist}
\usepackage{xspace}
\usepackage[lofdepth,lotdepth]{subfig}

\newcommand{\cellbg}[1]{\cellcolor{lightgray}\textbf{#1}}
\usepackage{colortbl}

\definecolor{orangeX}{rgb}{1,.5,0}
\definecolor{blueX}{rgb}{.2, .59, .88}
\definecolor{purpleX}{rgb}{.294118, 0, .509804}
\definecolor{greenX}{rgb}{.421, .578, .241}
\definecolor{bole}{rgb}{0.47, 0.27, 0.23}
\definecolor{mypink3}{cmyk}{0, 0.7808, 0.4429, 0.1412}
\definecolor{mygray}{gray}{0.6}
	
\newcommand{\audeep}{\mbox{\textsc{auDeep}}}

\newcommand{\ds}{\textsc{DeepSpectrum}}

\newcommand{\eg}{e.\,g., }

\copyrightyear{2022}
\acmYear{2022}
\setcopyright{acmlicensed}\acmConference[ComParE '22]{Proceedings of the ACM Multimedia}{October 22}{Lisbon, Portugal}
\acmBooktitle{Proceedings of the ACM Multimedia (MM '22), October, 2022, Lisboa, Portugal}
\acmPrice{15.00}
\acmDOI{10.1145/3475957.3484450}
\acmISBN{978-1-4503-8678-4/21/10}
\begin{document}
\fancyhead{}
\title{
The ACM Multimedia 2022 Computational Paralinguistics Challenge: Vocalisations, Stuttering, Activity, \& Mosquitoes} 

\author{Bj\"orn W.\ Schuller}
\affiliation{%
  \institution{Imperial College London}
  \city{London, United Kingdom}}
\renewcommand{\shortauthors}{Schuller, et al.}

\author{Anton Batliner}
\affiliation{%
  \institution{University of Augsburg}
  \city{Augsburg, Germany}}

\author{Shahin Amiriparian}
\affiliation{%
  \institution{University of Augsburg}
  \city{Augsburg, Germany}}

\author{Christian Bergler}
\affiliation{%
  \institution{FAU}
  \city{Erlangen-Nuremberg, Germany}}
  
  \author{Maurice Gerczuk}
\affiliation{%
  \institution{University of Augsburg}
  \city{Augsburg, Germany}}

\author{Natalie Holz}
\affiliation{%
  \institution{MPI}
  \city{Frankfurt, Germany}}

\author{Pauline Larrouy-Maestri}
\affiliation{%
  \institution{MPI}
  \city{Frankfurt, Germany}}

\author{Sebastian P.\ Bayerl}
\affiliation{%
  \institution{TH N\"urnberg}
  \city{N\"urnberg, Germany}}
  
\author{Korbinian Riedhammer}
\affiliation{%
  \institution{TH N\"urnberg}
  \city{N\"urnberg, Germany}}
  
\author{Adria Mallol-Ragolta}
\affiliation{%
  \institution{University of Augsburg}
  \city{Augsburg, Germany}}
  
\author{Maria Pateraki}
\affiliation{%
  \institution{FORTH}
  \city{Heraklion, Greece}}

\author{Harry Coppock}
\affiliation{%
  \institution{Imperial College London}
  \city{London, United Kingdom}}
  
\author{Ivan Kiskin}
\affiliation{%
  \institution{University of Surrey}
  \city{Guildford, United Kingdom}}

\author{Marianne Sinka}
\affiliation{%
  \institution{University of Oxford}
  \city{Oxford, United Kingdom}}
  
\author{Stephen Roberts}
\affiliation{%
  \institution{University of Oxford}
  \city{Oxford, United Kingdom}}

\begin{abstract}
The ACM Multimedia 2022 Computational Paralinguistics Challenge addresses four different problems for the first time in a research competition under well-defined conditions: 
In the \textit{Vocalisations} and \textit{Stuttering} Sub-Challenges, a classification on human non-verbal vocalisations and speech has to be made;
the \textit{Activity} Sub-Challenge aims at beyond-audio human activity recognition from smartwatch sensor data; and 
in the \textit{Mosquitoes} Sub-Challenge, mosquitoes need to be detected.
We describe the Sub-Challenges, baseline feature extraction, and classifiers based on the `usual' \textsc{ComParE} and BoAW features,
the \audeep{} toolkit, 
and deep feature extraction from pre-trained CNNs using the \ds{} toolkit; in addition, we add end-to-end sequential modelling, 
and a \textsc{log-mel-128-BNN}.
\end{abstract}



\maketitle
\section{Introduction}

	In this ACM Multimedia 2022 \textsc{COMputational PARalinguistics  challengE  (\textsc{ComParE})} -- the 14th since 2009 \cite{Schuller11-RRE},
	we address four new problems within the field of Computational Paralinguistics \cite{Schuller14-CPE} in a challenge setting:


In the \textbf{Vocalisations} Sub-Challenge,
non-verbal vocal expressions from the Variably Intense Vocalizations of Affect and Emotion Corpus \cite{holz2022variably,holz2021paradoxical} are used (\textbf{VOC-C}) for classifying the expression of six different emotions. Such human non-verbals are still understudied 
but are ubiquitous in human communication 
\cite{pisanski2022form}.



In the \textbf{Stuttering} Sub-Challenge, parts (\textbf{KSF-C}) of the Kassel State of Fluency corpus  \cite{Bayerl20-TAA,Bayerl22-KTK}  
are used.
Stuttering is a complex speech disorder with a crude prevalence of about 1\,\% of the population \cite{Sommer21-PAT}.
Monitoring of stuttering would allow objective feedback to persons who 
stutter (PWS) and speech therapists, thus facilitating tailored speech therapy, with the automatic detection of different stuttering phenomena as a necessary prerequisite.

The human activity recognition corpus harAGE as used in the \textbf{Activity} Sub-Challenge (\textbf{HAR-C}), provided by the EU Horizon 2020 project sustAGE \cite{Mallol-Ragolta22-S1F}, is a multimodal dataset 
collected with the smartwatch Garmin Vivoactive 3 \cite{Mallol21-HAN,Mallol22-OPF}. 
 The monitoring of different types of physical activity
vs inactivity 
is of vital importance to promote healthier and active life styles in the population, improving their overall physical health and wellbeing \cite{Fox99-TIO,Penedo05-EAW}. 


The Mosquito corpus as used in the \textbf{Mosquitoes} Sub-Challenge (\textbf{MOS-C}), provided by the HumBug Project, is a large-scale audio database consisting of over 20\,hours of mosquito flight recordings (HumBugDB \cite{Kiskin21-HAL}). Mosquitoes are responsible for more human deaths than any other creature; \eg in 2020 malaria caused around 241 million cases of disease across more than 100 countries resulting in an estimated 627\,000 deaths \cite{world2021world}. %
It is imperative therefore to accurately locate and identify dangerous mosquitoes to achieve efficient mosquito control.

	
For all tasks, a target class has to be predicted for each case. 
Contributors can employ their own features and machine learning (ML) algorithms; standard feature sets and procedures are provided.
Participants have to use the pre-defined
partitions for each Sub-Challenge. 
They may report results that they 
obtain from the \textbf{Train}(ing)/\textbf{Dev}(elopment) 
set 
but have only 
five trials to upload their results on the \textbf{Test} set per Sub-Challenge, whose labels are unknown to them. 	
Each participation must be accompanied by a paper presenting the results, which undergoes peer-review.
The organisers preserve the right to re-evaluate the findings, but will not participate in the Challenge. 
	As evaluation measure, 
    we employ for all Sub-Challenges but Mosquitoes the \textbf{Unweighted Average Recall (UAR)} as used since the first Challenge from 2009 \cite{Schuller09-TI2,Schuller11-RRE};
    it is more adequate for (unbalanced) multi-class classifications than Weighted Average Recall (i.\,e., accuracy)
      \cite{Schuller14-CPE}.
      The Mosquitoes Sub-Challenge is an audio event detection task; 
      hence, we utilise 
      the \textbf{Polyphonic Sound Event Detection Score (PSDS)} \cite{bilen2020framework} 
      -- an extension for a classifier threshold-independent event-based $F$-score.
	Ethical approval for the studies has been obtained. 
  %

	\section{The Four Sub-Challenges}
	\label{Corpora}


\noindent
\textbf{Vocalisations -- The \textbf{Vocalisation} Corpus \textbf{VOC-C}: }\\
It is provided by the MPI for Empirical Aesthetics, Frankfurt am Main, featuring vocalisations -- such as laughter, cries, moans, or screams -- with different affective intensities, expressing different emotional states. The data from the female speakers have been made available 
to the public, see \cite{holz2021paradoxical, holz2022variably}; the male speakers are so far unseen. 
We partition the female vocalisations into Train (6 speakers, 625 samples) and Dev(elopment) (5 speakers, 460 samples), and the male vocalisations (2 speakers, 276 samples) into  Test, modelling a 6-class problem with the emotional classes \textit{achievement, anger, fear, pain, pleasure,} and \textit{surprise}. 
  
%

\noindent
\textbf{Stuttering -- The Kassel State of Fluency Corpus \textbf{KSF-C}:}  \\
The corpus provided by the TH N\"urnberg and the Kasseler Stottertherapie is derived from the Kassel State of Fluency (KSoF) corpus \cite{Bayerl20-TAA,Bayerl22-KTK}.
The original corpus features 5\,597 typical and nontypical (stuttering) 
3\,s 
segments from 37 German speakers with an overall duration of 
4.6\,h. 
The segments were annotated by three labellers as one of 7 classes (\textit{block, prolongation, sound repetition, word/phrase repetition, modified speech technique, interjection, no disfluency}) and with some additional information, e.\,g., about the recording quality.
Annotators were able to assign more than one label per segment.
For this challenge, we removed all the ambiguously labelled segments, thus only featuring 4\,601 segments.
The task proposed in this challenge is the classification of speech segments as one of 8 classes -- the seven stuttering-related classes and an eighth \textit{garbage} class, denoting segments that are unintelligible, contain no speech, or are negatively affected by background noise.
The dataset is split 
into three speaker-independent partitions 
(Train: 23 speakers, Dev: 6 speakers, Test: 8 speakers).


\noindent
\textbf{Activity -- The Human Activity Recognition Corpus HAR-C:} \\
The harAGE corpus\footnote{\url{https://zenodo.org/record/6517688}} \cite{Mallol21-HAN,Mallol22-OPF} contains 
17\,h 37\,m 20\,s of triaxial accelerometer, heart rate, and pedometer sensor measurements
from 30 (14\,f, 16\,m) participants with a mean age of 40.0\,years and a standard deviation of 8.3\,years. 
%
Sensor measurements from 
eight activities are included: \textit{lying}, \textit{sitting}, \textit{standing}, \textit{washing hands}, \textit{walking}, \textit{running}, \textit{stairs climbing}, and \textit{cycling}. The dataset is split into three participant-independent and gender-balanced partitions. The Train, Dev, and Test partitions contain a total of 10\,h 41\,m 20\,s, 2\,h 16\,m 0\,s, and 4\,h 40\,m 0\,s of data from 17 (8\,f, 9\,m), 6 (3\,f, 3\,m), and 7 (3\,f, 4\,m) participants, respectively. 
Each sample in the harAGE corpus contains 20\,s of continuous sensor measurements from one participant performing one of the different activities considered in the dataset. The task in this Sub-challenge consists in the development of unimodal and/or multimodal systems able to analyse these 20\,s of sensor measurements and infer the corresponding activity.


\noindent
\textbf{Mosquitoes -- The Mosquito Corpus \textbf{MOS-C}:} \\
It is provided by the HumBug Project\footnote{The full list of authors contributing to HumBugDB is in \cite{Kiskin21-HAL} and associated Zenodo repository} and is strongly based on HumBugDB \cite{Kiskin21-HAL}. In a revision for this Sub-challenge\footnote{{\texttt{v0.0.2 HumBugDB:}} \url{https://zenodo.org/record/6478589}}, the former test set A is expanded with more challenging negatives and now forms Dev A. The former test set B forms Dev B, and the training set is identical. The task is to detect \textit{timestamps for acoustic mosquito events} -- Mosquito Event Detection (MED). The challenge is therefore scored in the time domain with the PSDS package \cite{bilen2020framework}. Details of 
Train and Dev 
are given 
in \cite[Sec.\ 4]{Kiskin21-HAL}. To summarise, Dev A represents semi-field conditions, where mosquitoes were manually released near recording setups that feature traditional housing constructions, equipped with mosquito bednets \cite[Sec.\ 2.1.2]{sinka2021humbug}. Dev B is a low-SNR recording set of free-flying mosquitoes within culture cages. The data vary in sample rate, recording devices, ambient conditions, and experimental assumptions. As these factors can introduce confounding, they are given as metadata and documented in \cite[Appx.\ C]{Kiskin21-HAL}. 
The test set consists of recordings conducted in South East Tanzania by volunteers in people's homes. It is therefore not included in the hosted Zenodo dataset due to the sensitive nature of the data. Please note that participants will not receive the test data but will submit dockerised versions of their code using the help of the provided templates for either Tensorflow 2.0 \cite{tensorflow2015-whitepaper}, or PyTorch \cite{NEURIPS2019_9015}, that participants are free to choose as they wish.

\begin{table*}[t!]
\caption{Summary of the databases presented per Sub-Challenge. Number of instances per class in the Train/Dev/Test splits. 
The test split distributions are blinded during the ongoing challenge and will be given in the final version. 
}
\label{tab:db}
\centering
\resizebox{\textwidth}{!}{
\begin{tabular}{lrrrr|lrrrr|lrrrr|lrrrr} \toprule
\multicolumn{5}{c|}{\textbf{VOC-C}: classification task (\#)}  & \multicolumn{5}{c|}{\textbf{KSF-C}: classification task (\#)} 
& \multicolumn{5}{c|}{\textbf{HAR-C}: classification task (\#)} &   \multicolumn{5}{c}{\textbf{MOS-C}:  detection (in hours)}   \\ \midrule
Class  &  Train & Dev & Test & $\Sigma$  &   Class &  Train & Dev & Test & $\Sigma$ &   Class &  Train & Dev  & Test & $\Sigma$  & Class &  Train & Dev A/B & Test & $\Sigma$\\ \midrule
achiev.  & 89    & 72    & --    & --   &     Block        & 310   & 102   & -- & -- &  lying        &  257  &  61  & --  & --   &  mosquito &17.0 & 1.1/0.25 & -- & -- \\ 
anger         & 101   & 73    & --    & --   &    Fillers      & 205   & 104   & --& -- &  sitting      &  238  &  57  & --  & --    &  non-mosquito & 13.4 &  2.7/0.56 & --  & -- \\
fear          & 103   & 73    & --    & --   &   Garbage      & 52    & 33    &-- & -- &  standing     &  244  &  57  & --  &  --     & & & & &\\   
pain          & 114   & 71    & --    & --   &    Modified     & 687   & 185   &-- & -- &  wash. hands&  133  &  35  & --  & --      & & & & & \\
pleasure      & 109   & 93    & --    & --   &   Prolong. & 183   & 53    & --& -- & walking      &  302  &  57  & --  & --  & & & & &\\
surprise      & 109   & 78    & --    & --   &    SoundRep.    & 169   & 38    & --& -- &  running      &  301  &  40  & --  & -- &  & & & & \\
              &       &       &       &     &  WordRep.     & 53    & 23    & --& -- &  stairs climb.&  263  &  43  & --  & -- & & & & & \\
              &       &       &       &      &  no\_disfl.   & 830   & 444   & --& -- &  cycling      &  186  &  58  & --  & --  & & & & &  \\   \midrule
$\Sigma$      & 625   & 460   & 276   & 1\,361 &    $\Sigma$    & 2\,489 & 982   & 1\,130 &  4\,601  &  $\Sigma$     & 1\,924  & 408  & 840  & 3\,172 &  $\Sigma$  & 30.4 & 3.8/0.81 & 18 & 53.01\\
\bottomrule 
\end{tabular}
}
\end{table*}

\section{Experiments and Results}
\label{experiments}
\vspace{-0.1cm}

For the VOC-C and the KSF-C, 
the segmented 
audio was converted to single-channel 16\,kHz, 16\,bits PCM format.
Table \ref{tab:db} shows the number of 
data for Train, Dev, and Test for the different 
corpora. 
MOS-C Dev 
was split into two sets with differing conditions.

\begin{table*}[t!]
    \caption{Results for the Sub-Challenges. The \textbf{official baselines} for Test are highlighted (bold and greyscale); 
        there are \textbf{no} official baselines for Dev.  \textbf{UAR}: Unweighted Average Recall. 
	    CI on Test: Confidence Intervals on Test, see explanation in the text. 
	    	    }
    \centering
    \begin{footnotesize}
    \begin{tabular}{lccc|ccc|lccc|lccc}
    \toprule
    \% & \multicolumn{3}{c}{Vocalisations: \textbf{VOC-C}: UAR} & \multicolumn{3}{c}{Stuttering: \textbf{KSF-C}: UAR} & & \multicolumn{3}{c}{Activity: \textbf{HAR-C}: UAR} & & \multicolumn{3}{c}{Mosquitoes: \textbf{MOS-C}: PSDS}  \\
    \midrule
   Approach  & Dev & Test & CI on Test & Dev & Test & CI on Test & Approach  & Dev & Test & CI on Test &   Approach  &   Dev A/B & Test & CI on Test    \\      \midrule
    ComParE    & 39.8 & 32.7 & 27.7 -- 38.0    & 30.2 & 37.6 & 33.5 -- 41.4            &    HR               & 34.4 & 30.2 & 28.0 -- 32.3    &    mel-BNN  &    61.4/3.4 & \cellbg{6.4} & 6.0 -- 6.9 \\
    DeepSpectrum   & 35.0 & 34.1 & 29.5 -- 39.2   & 28.1 & \cellbg{40.4} & 36.4 -- 44.2  &  Steps             & 36.4 & 32.1 & 30.6 -- 33.7     &   &  \\
    auDeep        & 31.0  &  31.2 & 26.1 -- 36.6    & 17.7 & 25.9 & 21.9 -- 30.3          &    XYZ                & 65.8 & 69.3 & 66.6 -- 72.2 &   &      \\
   BoAWs     & 39.6   & \cellbg{37.4} & 32.6 -- 42.8    & 26.7 & 32.1 & 28.2 -- 36.0 &   HR$\oplus$Steps       & 52.2 & 42.1 & 39.6 -- 44.5 &   &   \\
    Fusion          & 39.8     & 36.1 & 31.3 -- 31.3    & 28.7 & 38.3 & 34.3 -- 41.9     &  HR$\oplus$XYZ          & 74.5 & 63.9 & 61.0 -- 66.9   &  &    \\
             &&&&&&                                                         &    Steps$\oplus$XYZ            & 66.6 & 65.5 & 62.5 -- 68.4     &    &   \\  
             &&&&&&                                                         &     HR$\oplus$Steps$\oplus$XYZ & 77.7 & \cellbg{72.2} & 69.2 -- 75.0  &   &       \\           
    \bottomrule
    \end{tabular}
    \label{tab:resultsMerged}
    \end{footnotesize}
\end{table*}

	\subsection{Approaches}
	\label{ssec:approaches}


\noindent
\textbf{\textsc{ComParE} Acoustic Feature Set: }
The official baseline feature set from openSMILE is the same as has been used in previous editions of the \textsc{ComParE} challenges, starting from 2013~\cite{Schuller13-TI2}.
It is described in 	\cite{Eyben13-RDI,Schuller13-TI2}.
	
	%

\noindent\textbf{\ds{}:}\footnote{\url{https://github.com/DeepSpectrum/DeepSpectrum}} 
It is applied to obtain deep representations from the input audio data utilising image pre-trained Convolutional Neural Networks (CNNs)~\cite{Amiriparian17-SSC}. 
It has been used in previous challenges \cite{schuller2020interspeech,schuller2021interspeech} and is described in  \cite{Amiriparian17-SSC}. A lightweight version of \ds{} for 
audio signal processing on-device 
can be found in~\cite{Amiriparian22-DAP}\footnote{\url{https://github.com/DeepSpectrum/DeepSpectrumLite}}.



\noindent\textbf{\audeep{}:}\footnote{\url{https://github.com/auDeep/auDeep}}
This feature set is obtained through unsupervised representation learning with recurrent sequence-to-sequence autoencoders
\cite{Amiriparian17-STS,Freitag18-AUL}; it has as well been employed in previous challenges \cite{schuller2020interspeech,schuller2021interspeech}.
Learnt representations of a spectrogram are extracted and then concatenated to obtain a final feature vector.


  \noindent  \textbf{Bag-of-Audio-Words (BoAWs):}
Audio chunks are represented as histograms of ComParE LLDs, after quantisation based on a codebook. They have been used in previous challenges \cite{schuller2020interspeech,schuller2021interspeech} and other studies   \cite{Lim15-RSE,Amiriparian18-BND,Schmitt16-ATB}; the toolkit openXBOW is
described in \cite{Schmitt17-OIT}.


\begin{figure*}[t!]
    \subfloat[VOC-C Confusion Matrix]{\includegraphics[width=0.32\linewidth]{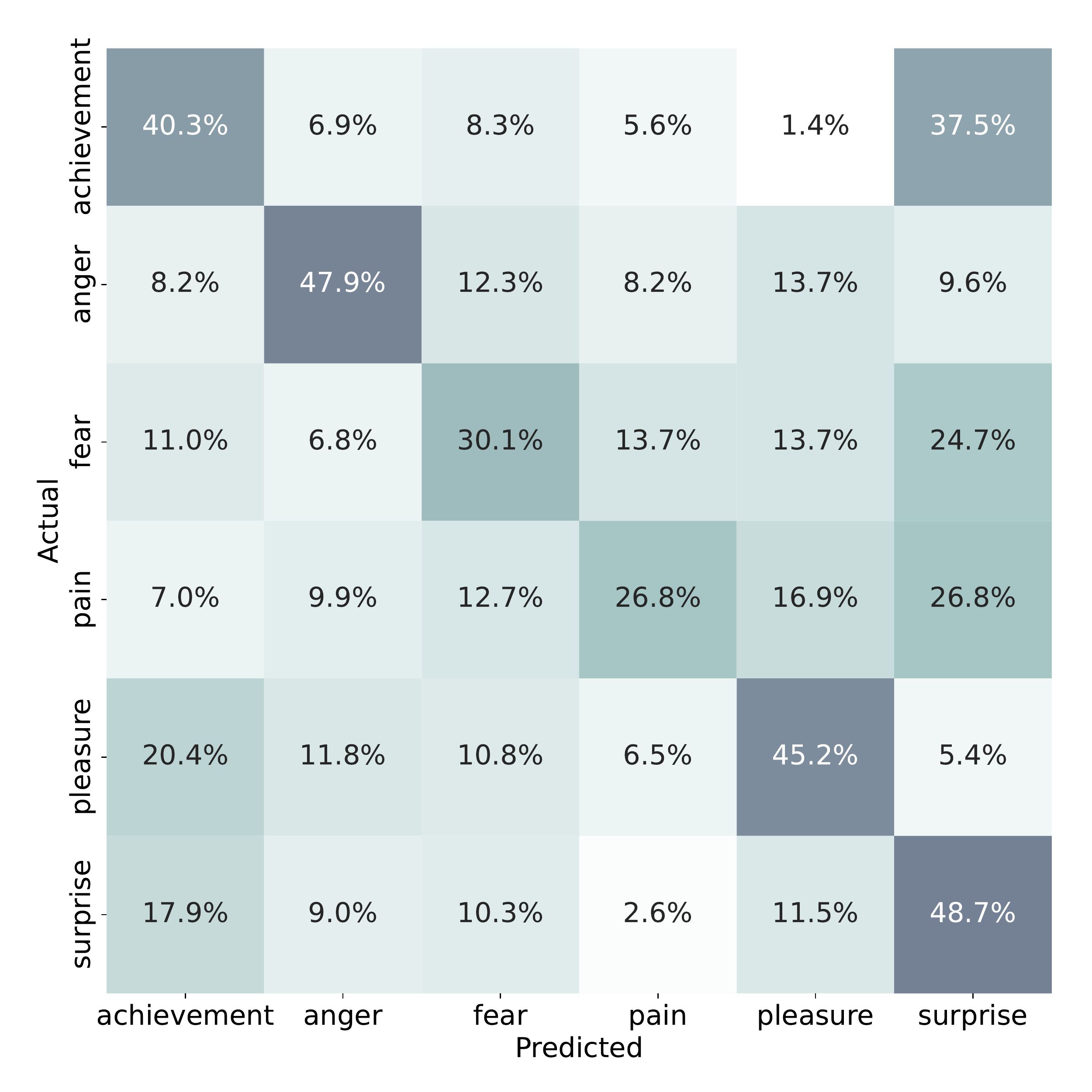}}
    \label{fig:voc-c-cm}
    \subfloat[KSF-C Confusion Matrix]{
    \includegraphics[width=0.32\linewidth]{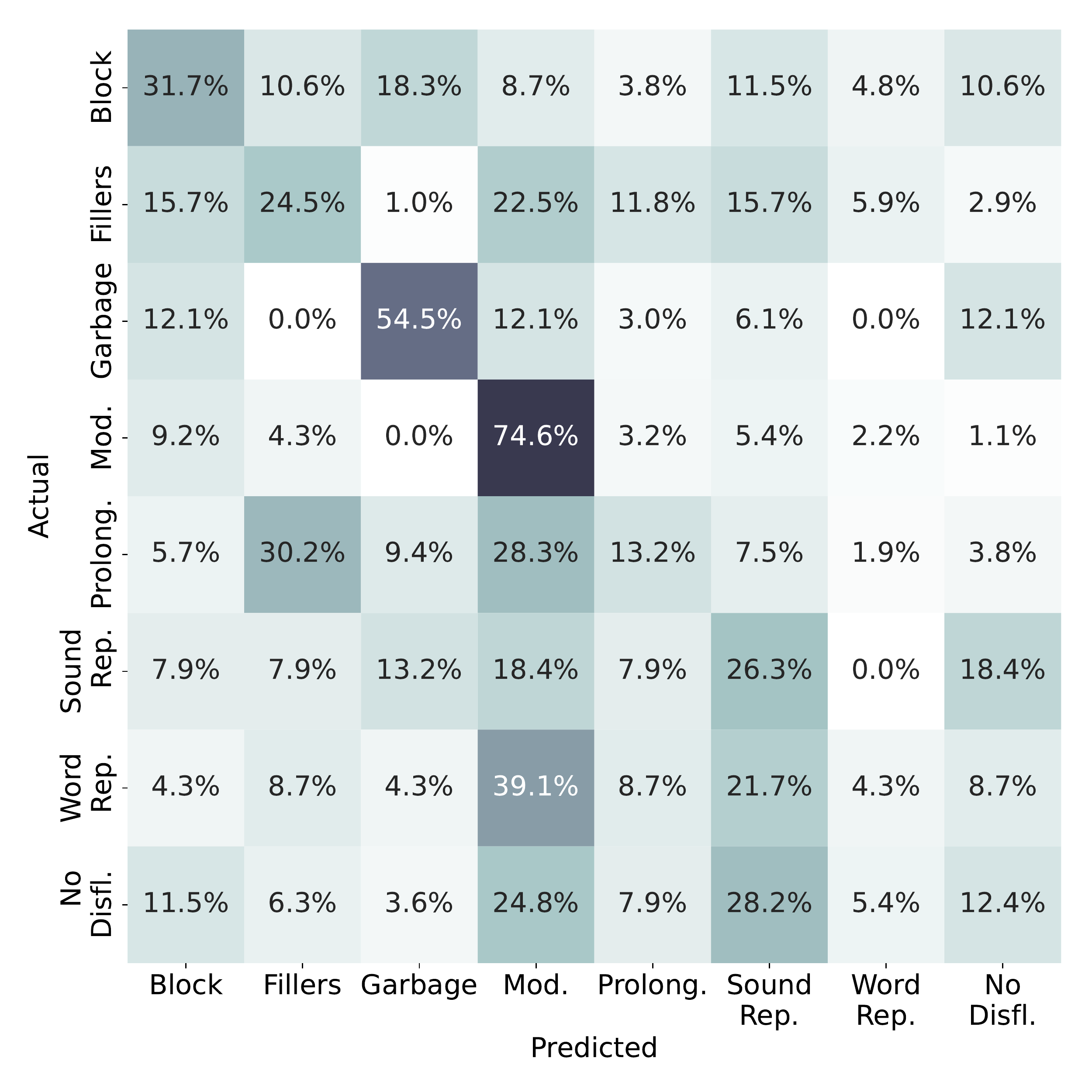}}
    \label{fig:ksf-c-cm}
    \subfloat[HAR-C Confusion Matrix]{\includegraphics[width=0.32\linewidth]{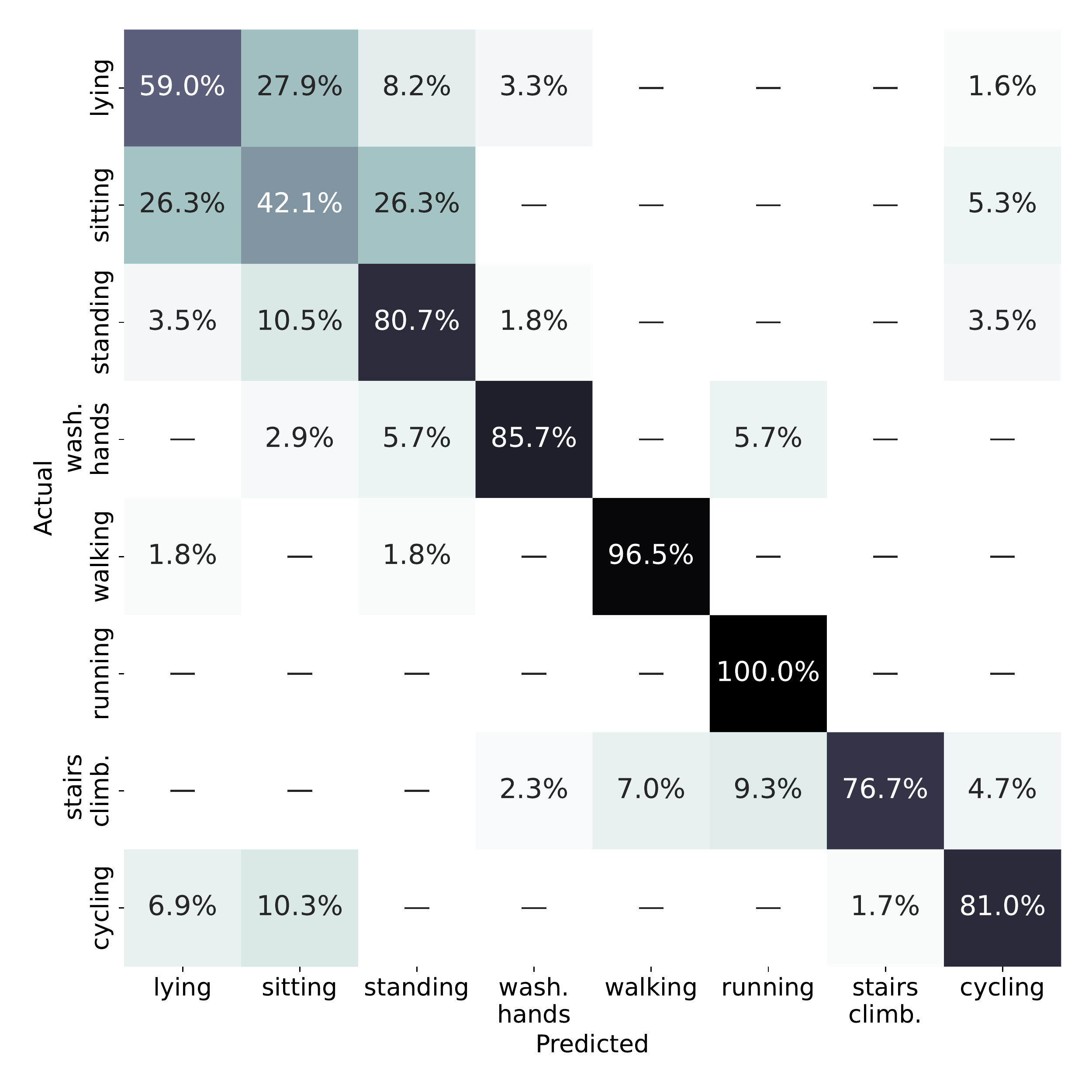}}
    \label{fig:har-c-cm}
\caption{
Confusion matrices for \textbf{VOC-C}, \textbf{KSF-C}, and \textbf{HAR-C} on Dev. 
The individual approach/hyperparameters performing best on Dev (without fusion) are chosen; see Table \ref{tab:resultsMerged}.      
In the cells, 
the percent of `classified as' of the class displayed in the respective row are given; percentage also indicated by colour-scale: the darker, the higher. Cases per class are given in Table \ref{tab:db}.}
\label{fig:dev}
\end{figure*}

\noindent \textbf{End-to-end sequential modelling:} 
The HAR-C implements an end-to-end approach exploiting the sensor data as input. As described in \cite{Mallol22-OPF}, 3-dimensional, 2-dimensional, and 9-dimensional traces are generated from the raw heart rate, pedometer, and triaxial accelerometer measurements, respectively. As opposed to \cite{Mallol22-OPF}, herein, we do not debias the accelerometer measurements to ease the deployment of the presented approach in real-life applications. The network implemented is composed of a dedicated feature extraction block for each modality -- responsible for extracting deep learnt representations from the input traces -- followed by a classification block -- in charge of performing the actual inference. 
The feature extraction block implements a 1-dimensional convolutional layer, and the classification layer two fully connected layers. 
The dimensionality of the resulting features at the output of the feature extraction block depends on the number of modalities to be fused, concatenating the embedded representations learnt separately from each modality. 

\noindent
{\textbf{\textsc{log-mel-128-BNN}:} MOS-C utilises a Bayesian Convolutional Neural Network with four convolutional, two max-pooling, and one fully connected layer augmented with dropout layers \cite[Appx.\ B.4]{Kiskin21-HAL}. Its structure is based on prior models that have been successful in assisting domain experts for mosquito tagging \cite{kiskin2021automatic}. 
As features, the baseline uses 128 log-mel spectrogram coefficients with a time window of 30 feature frames and a stride of 5 frames for training. Each frame spans 64\,ms, forming a single training example $\mathbf{X}_{i} \in \mathbb{R}^{128\times30}$ with a temporal window of 1.92\,s. Dev and Test events may, however, be scored on shorter time windows.}

\vspace{-0.1cm}
\subsection{Challenge Baselines and Interpretation}
\label{challenge_baselines}

We provide a branch on the official challenge repository\footnote{\url{https://github.com/EIHW/ComParE2022}} for each Sub-Challenge, which includes scripts allowing participants to fully reproduce the baselines (including pre-processing, model training, and model evaluation on Dev). 
For VOC-C, KSF-C, and HAR-C, the 95\,\% Confidence Intervals (CI) were computed by $1\,000$x bootstrapping (random sampling with replacement) and UARs for Test, based on the same model that was trained with Train and Dev.
For MOS-C, as appropriate for sound event detection tasks, the 95\,\% CI intervals are constructed using the jackknife method (leave-one-out sampling) \cite{jackknife} with the number of samples equal to the number of test audio recordings.
Due to space restrictions, for VOC-C and KSF-C, we leave out the results for every hyperparameter configuration evaluated and only provide the best results obtained.
%
The baselines for both VOC-C and KSF-C consist of using Support Vector Machines with linear kernels on four different audio feature representations -- ComParE, DeepSpectrum, auDeep, and BoAWs. 
All feature representations are scaled to zero mean and unit standard deviation,
 using the parameters from the respective training set (when Train and Dev are fused for the final classifier, the parameters are calculated on this fusion). 
The SVM complexity parameter $C$ is always optimised during the development phase.

\noindent
\textbf{Vocalisations -- VOC-C:}  
We obtain the best \textbf{UAR=37.4\,\%} on Test with BoAWs, see Table \ref{tab:resultsMerged}.  
Figure \ref{fig:dev}(a) shows, for the best Dev result
given in Table  
\ref{tab:resultsMerged}, that the two classes \textit{pain} and \textit{fear} fall behind the other four classes, and that they are mostly confused with surprise. 
%

\noindent
\textbf{Stuttering -- KSF-C:} We achieve \textbf{UAR=40.4\,\%} on Test with DeepSpectrum. Looking at the confusion matrix of our best result on Dev in Figure \ref{fig:dev}(b), word repetitions seem to be the hardest to detect and differentiate, especially from instances of modified speech and sound repetitions. 
\noindent 
\textbf{Activity -- HAR-C}: 
The best approach on Test fuses the heart rate, the pedometer, and the accelerometer modalities, scoring a \textbf{UAR=72.2\,\%}. The results 
highlight the importance of the accelerometer information for this task, as the 
models exploiting this modality outperform those using the heart rate and the pedometer information, either unimodally or multimodally. 
Analysing the confusion matrix given in Figure \ref{fig:dev}(c) for the best result on Dev in Table \ref{tab:resultsMerged}, we observe that the main confusions take place among  
the `non-moving' activities lying, sitting, and standing. 

\noindent
\textbf{Mosquitoes -- MOS-C}:
Table \ref{tab:resultsMerged} shows the 
PSDS of 61.4\,\% and 3.4\,\% 
achieved 
on Dev A and B, respectively. We note that 
the provided model is unable to achieve a good score on Dev B, which features a lower SNR, more challenging dataset. The baseline scores \textbf{6.4\,\%  PSDS} on the Test partition, which can be thought of as an approximate combination in recording conditions of Dev A and B. The results highlight the need to train a model that is able to perform well and generalise across different deployment scenarios. Each of the Dev A, Dev B, and Test sets features considerably different audio backgrounds, as they are recorded in different environments. Additional \textit{feature window-based} metrics are supplied in the repository, which give a breakdown of performance by precision-recall, ROC, and confusion matrices. These may be helpful for developing with the ultimate aim of maximising the PSDS on Test.

\vspace{-0.2cm}
\section{Concluding Remarks}
\label{Conclusion}

This year's challenge is new by four new tasks, all of them highly relevant for applications. 
We feature our `classic' approaches 
\textbf{ \textsc{ComParE}} and 
\textbf{Bag-of-Audio-Words (BoAWs)},
\textbf{ \textsc{auDeep} }, and 
\textbf{ \ds{}} for VOC-C and KSF-C, 
and two new ones, tailored for HAR-C and MOS-C.
For all computation steps, scripts are provided that can, but need not be used by the participants.
We expect participants to obtain better performance measures by employing novel (combinations of) procedures and features, including such tailored to the particular tasks.

\section{Acknowledgments}
We acknowledge funding from the European Union's Horizon 2020 research and innovation programme under grant agreement No.\ 826506 (sustAGE), from the Deutsche Forschungsgemeinschaft (DFG) under grant agreement No.\ 421613952 (ParaStiChaD), from 
the DFG's Reinhart Koselleck project No.\ 442218748 (AUDI0NOMOUS), and from the Gates Foundation No.\ opp1209888, as well as the contributions of all authors in \cite[HumBugDB]{Kiskin21-HAL} and MOS-C Zenodo repository.

\begin{acronym}
\acro{AReLU}[AReLU]{Attention-based Rectified Linear Unit}
\acro{AUC}[AUC]{Area Under the Curve}
\acro{CCC}[CCC]{Concordance Correlation Coefficient}
\acro{CNN}[CNN]{Convolutional Neural Network}
\acrodefplural{CNN}[CNNs]{Convolutional Neural Networks}
\acro{CI}[CI]{Confidence Interval}
\acrodefplural{CI}[CIs]{Confidence Intervals}
\acro{CCS}[CCS]{COVID-19 Cough}
\acro{CSS}[CSS]{COVID-19 Speech}
\acro{CTW}[CTW]{Canonical Time Warping}
\acro{ComParE}[ComParE]{Computational Paralinguistics Challenge}
\acrodefplural{ComParE}[ComParE]{Computational Paralinguistics Challenges}
\acro{DNN}[DNN]{Deep Neural Network}
\acrodefplural{DNNs}[DNNs]{Deep Neural Networks}
\acro{DEMoS}[DEMoS]{Database of Elicited Mood in Speech}
\acro{eGeMAPS}[\textsc{eGeMAPS}]{extended Geneva Minimalistic Acoustic Parameter Set}
\acro{EULA}[EULA]{End User License Agreement}
\acro{EWE}[EWE]{Evaluator Weighted Estimator}
\acro{FLOP}[FLOP]{Floating Point Operation}
\acrodefplural{FLOP}[FLOPs]{Floating Point Operations}
\acro{FAU}[FAU]{Facial Action Unit}
\acrodefplural{FAU}[FAUs]{Facial Action Units}
\acro{GDPR}[GDPR]{General Data Protection Regulation}
\acro{HDF}[HDF]{Hierarchical Data Format}
\acro{Hume-Reaction}[\textsc{Hume-Reaction}]{Hume-Reaction}
\acro{HSQ}[HSQ]{Humor Style Questionnaire}
\acro{IEMOCAP}[IEMOCAP]{Interactive Emotional Dyadic Motion Capture}
\acro{KSS}[KSS]{Karolinska Sleepiness Scale}
\acro{LIME}[LIME]{Local Interpretable Model-agnostic Explanations}
\acro{LLD}[LLD]{Low-Level Descriptor}
\acrodefplural{LLD}[LLDs]{Low-Level Descriptors}
\acro{LSTM}[LSTM]{Long Short-Term Memory}
\acro{MIP}[MIP]{Mood Induction Procedure}
\acro{MIP}[MIPs]{Mood Induction Procedures}
\acro{MLP}[MLP]{Multilayer Perceptron}
\acrodefplural{MLP}[MLPs]{Multilayer Perceptrons}
\acro{MPSSC}[MPSSC]{Munich-Passau Snore Sound Corpus}
\acro{MTCNN}[MTCNN]{Multi-task Cascaded Convolutional Networks}
\acro{SER}[SER]{Speech Emotion Recognition}
\acro{SHAP}[SHAP]{SHapley Additive exPlanations}
\acro{STFT}[STFT]{Short-Time Fourier Transform}
\acrodefplural{STFT}[STFTs]{Short-Time Fourier Transforms}
\acro{SVM}[SVM]{Support Vector Machine}
\acro{TF}[TF]{TensorFlow}
\acro{TSST}[TSST]{Trier Social Stress Test}
\acro{TNR}[TNR]{True Negative Rate}
\acro{TPR}[TPR]{True Positive Rate}
\acro{UAR}[UAR]{Unweighted Average Recall}
\acro{Ulm-TSST}[\textsc{Ulm-TSST}]{Ulm-Trier Social Stress Test}
\acrodefplural{UAR}[UARs]{Unweighted Average Recall}
\end{acronym}

\clearpage
\footnotesize
\bibliographystyle{ACM-Reference-Format}
\bibliography{sample-base}

\end{document}